# Position paper From the digital twins in healthcare to the Virtual Human Twin: a moon-shot project for digital health research

Marco Viceconti, Maarten De Vos, Sabato Mellone, and Liesbet Geris

***Abstract*—The idea of a systematic digital representation of the entire known human pathophysiology, which we could call the Virtual Human Twin, has been around for decades. To date, most research groups focused instead on developing highly specialised, highly focused patient-specific models able to predict specific quantities of clinical relevance. While it has facilitated harvesting the low-hanging fruits, this narrow focus is, in the long run, leaving some significant challenges that slow the adoption of digital twins in healthcare. This position paper lays the conceptual foundations for developing the Virtual Human Twin (VHT). The VHT is intended as a distributed and collaborative infrastructure, a collection of technologies and resources (data, models) that enable it, and a collection of Standard Operating Procedures (SOP) that regulate its use. The VHT infrastructure aims to facilitate academic researchers, public organisations, and the biomedical industry in developing and validating new digital twins in healthcare solutions with the possibility of integrating multiple resources if required by the specific context of use. Healthcare professionals and patients can also use the VHT infrastructure for clinical decision support or personalised health forecasting. As the European Commission launched the EDITH coordination and support action to develop a roadmap for the development of the Virtual Human Twin, this position paper is intended as a starting point for the consensus process and a call to arms for all stakeholders.**

## I. Introduction

EVEN though insights into the physiology of health and disease are increasing and there is a continuous development of new medical therapies, patients and healthcare professionals are confronted with unmet needs in all societies across the globe. Lack of adequate, accessible, and affordable treatment options, long waiting times and lack of attention for under-represented and under-served populations (e.g., paediatrics and orphan diseases) are some of the major issues reported by consumers and providers of healthcare services [1]. The increased incorporation of new technologies, such as computer modelling & simulation, in the R&D process and the organisation and provision of healthcare services could help alleviate the unmet needs.

The IUPS Physiome project [2] first introduced the idea of a systematic digital dynamic representation of the entire known human physiology, much like the Human Genome Project [3] provided a complete mapping of the human genome. The Virtual Physiological Human (VPH) initiative moved the focus of the research initiative to use such computer models as clinical decision support systems [4]. This shift dramatically increased the potential for clinical impact, focusing the application of these new technologies. Digital Twins in Healthcare (DTH) solutions (initially referred to as Digital Patient solutions [5]) developed into patient-specific computer models designed to accurately predict specific quantities that, while difficult or impossible to measure directly, would be very useful to support a specific clinical decision. These models use all available mechanistic knowledge on the disease progression and the underlying pathology to provide highly specialised clinical decision support tools. A perfect example of this first generation of technologies is HeartFlow FFRCT (HeartFlow Inc., USA), where a patient-specific computer model informed by medical imaging data replaces an invasive measurement essential to select the optimal treatment in coronary stenosis patients [6]. DTHs are used in clinical practice as patient-specific clinical decision support (DSS) systems. DSS systems

Submitted on Dec 27th, 2022. The European Commission supported this study through the H2020 project “In Silico World: Lowering barriers to ubiquitous adoption of In Silico Trials” (topic SC1-DTH-06-2020, grant ID 101016503). The content largely reflects the consensus process done as of the Horizon Europe Coordination & Support Action EDITH (ID 101083771).

Marco Viceconti is with the Department of Industrial Engineering, Alma Mater Studiorum - University of Bologna (IT) and with the Medical Technology Lab, IRCCS Istituto Ortopedico Rizzoli, Bologna (IT) (e-mail: marco.viceconti@unibo.it). Maarten De Vos is with ESAT – Stadius, KU Leuven (BE) and with Development and Regeneration, KU Leuven (BE)(e-mail: maarten.devos@kuleuven.be). Sabato Mellone is with the Department of Electrical, Electronic and Information Engineering, Alma Mater Studiorum - University of Bologna (IT) (e-mail: sabato.mellone@unibo.it). Liesbet Geris is with VPH Institute for Biomedical Integrative Research, with GIGA In Silico Medicine, University of Liege (BE), and with the Biomechanics Section, KU Leuven (BE) (e-mail: director@vph-institute.org).

are broadly adopted in all specialities of healthcare. DSS systems are developed based on traditional statistical methods or more recent Artificial Intelligence (AI) based approaches, where patterns are unravelled from large quantities of data. What makes the DTH unique is that it provides a model that is unique to the individual, whereas DSS systems currently in use are developed for a population.

Though some digital twin solutions have developed into full-blown commercial products widely adopted in clinical practice, their adoption is still limited compared to their potential. The DTHs now on the market targeted, to some extent, the low-hanging fruits. We are far from the digital twin revolution that road-mapping documents [7], [4], [8]–[10] and "evangelists" like some of the authors of this paper prophesied in the past 20 years. While digital twin results have been impressive across various industries (e.g., manufacturing, see [11]), digital twins in healthcare have found their way into clinical applications only in a limited number of cases. However, the momentum is picking up, as can be appreciated by the increasing number of activities (workshops, task forces, white papers, etc.) on digital twins organised by a wide range of stakeholders.

The development, introduction, and uptake of new technologies in healthcare (or society at large) does not happen in isolation; it requires the development of an entire ecosystem to address specific unmet needs. Ecosystems are dynamic and co-evolving communities of diverse actors that come together around shared interests to realise outcomes beyond any individual actor. In the case of digital twins for healthcare, these actors include academia, industry, citizens, patients, healthcare professionals, payers and buyers, regulators, and policymakers. Given the complexity of the technological development, the academic and industrial partners involved originate from various communities, including (but not limited to) computer modelling and simulation, artificial intelligence, wearables, sensors, and high-performance computing. All those communities might associate a different meaning to the digital twin concept depending on the background. Building an ecosystem accelerates learning and innovation across disciplines and sectors whilst identifying and addressing common challenges.

This position paper aims to provide a vision, a sort of manifesto, from which our community can start to develop the blueprint for the Virtual Human Twin.

We acknowledge that the vision in this position paper is still incomplete and, to some extent, immature. The paper aims to open the debate in the research community so that as many experts as possible can contribute to the consensus process driven by the EDITH support action[1] (see also the EDTH Roadmap outline[2]).

## II. Terminology

*In silico* medicine is an emerging sector in healthcare technologies, and like all immature fields, the terminology is still very fluid. Here we provide some definitions of terms we will use in the rest of the paper. While there is no ecosystem-wide consensus on some of these terms, these definitions should remove any ambiguity as to their use in the context of this paper.

***In silico* Medicine**: the term *in silico* means carried out in the computer; in contrast to *in vitro* (on the bench), *ex vivo* (outside the living organism), or *in vivo* (inside the living organism), *in silico* medicine encompasses the use of *in silico* technologies in all aspects of the prevention, diagnosis, follow-up, prognostic assessment, and treatment of diseases. These can be generic or individualised models. *In silico* models can be classified by the amount of data and/or knowledge required. As most models need both to some degree, this leads to a spectrum of models, with the extremes being predominantly data-driven models (such as Artificial Intelligence (AI)) on the one end and predominantly knowledge-driven models (such as physics-based, mechanistic models) on the other. It should be noted that both kinds of models contain tentative knowledge. But knowledge-driven models are based on hypotheses that have resisted extensive falsification attempts. In contrast, data-driven models rely on hypotheses that emerged as correlations within the data, which does not necessarily imply causation.

**Digital Twin in Healthcare (DTH)**: the Discipulus roadmap [5], published in 2013, proposed the term Digital Patient, first introduced in 2009 [12], as "a digital representation of the integration of the different patients-specific models for better prediction and treatment of diseases to provide patients with an affordable, personalised and predictive care". We define a predictive model to be subject- or patient-specific if there is an expectation that such predictions will be sufficiently accurate per each subject of the target population to support the decision process specific to the context of use of the model. But in the last years, this definition merged with the concept of *Digital Twin*, a "virtual representation that serves as the real-time digital counterpart of a physical object or process". While conceptually proposed by Michel Grieves in 2002 [11] in the context of manufacturing in Industry 4.0, its first recognised implementation was that of NASA for spacecraft simulation in 2010. In medicine, the concept was introduced much earlier than that; to date, there are 589 publications indexed in PubMed that contain the string "digital twin" in the title or abstract. The first appearance dates back to 1994, although, until 2019, it was used only in a few papers per year. Hence, the concept emerged in healthcare more or less concurrently with other industrial sectors. However, its clinical adoption has proceeded at a much slower pace.

The most important feature of a DTH is that it must model an individual; thus, population models are not DTH. The main difference from Grieves's definition is that only a minority of digital twins in healthcare described in the literature have a real-time component. While this is still being debated, many use the term even when no real-time update is involved. Here we will assume that DTHs can or cannot be driven in real-time by sensors, depending on the use case. But what makes it a twin is that it is informed by personal patient data at least once in its entire lifecycle.

[1] https://www.edith-csa.eu/

[2] https://zenodo.org/record/7796845

**Virtual Physiological Human (VPH)**: the VPH is a concept that started as an EU-led initiative that ran from 2005, when the term was first proposed, to the end of projects funded in the Seventh Framework Program (FP7), around 2015. The term was defined as a "methodological and technological framework that, once established, will enable collaborative investigation of the human body as a single complex system". But despite this broad definition, the initiative produced mainly what is called Digital Twins in Healthcare today, focusing on a single organ system. Through the activities of the international scientific society with the same name (e.g., the VPH institute, established in 2011), the term became synonymous with "*in silico* medicine" or "the application of computer modelling and simulation in all areas of healthcare".

***In silico* Trials (IST)**: the Avicenna roadmap [10] defines it as "individualised computer simulation used in the development or regulatory evaluation of a medicinal product, device, or intervention". The key word is *individualised*; an *in silico* trial is performed using hundreds or thousands of digital twin models, each referring to a specific physical or virtual patient. So, to develop *In silico* Trials, we first need to develop Digital Twins that predict outcomes of regulatory interest. In the regulatory context, where the word trial has a more specific meaning, these are usually called *In silico Methodologies*.

**Intended Use (IU)** and **Context of Use (CoU)**: when a Digital Twin in Healthcare is used as a clinical decision support system from a regulatory perspective, it is considered a medical device, namely a Software as Medical Device (SaMD). In that case, the model's medical purpose is the *Intended Use* of the medical device. In the context of *In silico* Trials, the purpose of the model is called *Context of Use*, defined as a complete and precise statement that describes the appropriate use of the new methodology and how the qualified methodology is applied in the development and regulatory review of a specific class of medical products.

**Knowledge-driven and data-driven models**: in general, predictive models are built using a combination of data and prior knowledge. However, some predictive models are built exclusively using data (statistical predictors, machine learning models, AI). We will refer to these models built predominantly using data as data-driven models. Conversely, models built purely using prior knowledge do not exist; however, we will refer to models built predominantly using knowledge as knowledge-driven models.

## III. Challenges

The *In silico* World Community of Practice (ISW_CoP) is a free online community open to anyone with professional or educational interests in *In silico* Medicine; requests to join can be submitted here: https://insilico.world/community/. As of June 2022, the community hosts over 500 students and experts working in academia, industry, regulatory agencies, Clinical Research Organisations, consulting firms, software developers, etc. The community hosts a variety of discussion groups and develops best practices around modelling and simulation in healthcare.

Various stakeholders provided anecdotal reports on why developing and adopting DTH solutions was difficult, yet a systematic analysis was only recently made available. This changed when the *In Silico* World Community of Practice was used to poll what the various represented stakeholders believed to be the main barriers to the adoption of knowledge-driven DTH technologies for a specific context of use (development and derisking of new medical products, the so-called *In silico* Trials). Six barriers were identified:

B1) **Lack of advanced models**: the scarcity of predictive models advanced and robust enough to aspire a use in the reduction, refinement or replacement of the clinical trials required for the certification of new medical products; challenges to support *in silico* model development; challenges in combining various *in silico* technologies, covering the entire spectrum from data-driven (black box) to knowledge-driven (white box). It is particularly challenging to develop the DTHs compared to other industries because it should be accurate for the individual. At the same time, it is known that (abnormal) human physiology can vary widely across people.

B2) **Lack of available representative data for development and independent validation**: lack of widely available collections of curated medical data specifically designed for the development (e.g., for Analytical Artificial Intelligence (AI) predictors) and the validation of *In silico* Trials (IST) technologies; lack of a priori knowledge which data is needed for a particular DTH; the challenge of data sharing. Need for more explainable AI predictors.

B3) **Lack of clear regulatory pathways**: the need for generalisation of the credibility assessment proposed in the ASME VV-40 to include also models used to evaluate pharmaceutical and combination products; lack of harmonisation of the VV-40 in the CEN and ISO standardisation framework; the need for "First in kind" qualification of *In Silico* Trials methods for the evaluation of drugs with the European Medicine Agency (EMA); need for exploration of new regulatory pathways for class III medical devices and pharmaceutical products that allow a significant reduction of human experimentation with *In Silico* Trials technologies, and the adoption of rigorous post-marketing surveillance that monitors the validity of the *in silico* predictions as the new product is adopted.

B4) **Poorly informed stakeholders**: lack of correct and effective information to all stakeholders (policymakers, patients, doctors, regulators, healthcare payers, Contract Research Organisations (CRO), technology providers, executives of biomedical companies, etc.) on the opportunities and the risks associated with the use of *In silico* Trials technologies; the need for development and application of a rigorous risk-benefit evaluation framework to assess the adoption of IST technologies, from the perspective of each stakeholder.

B5) **Poor scalability and efficiency**: insufficient scalability of most *In Silico* Trials models; lack of efficient computing of large-scale population models in easy-to-use, safe, and trusted environments.

B6) **Lack of trained workforce**: insufficient opportunities for training and re-training of personnel with the necessary technical skills on *In Silico* Trials, to work in industry, regulatory agencies, research hospitals, etc.

While this analysis focused on *In Silico* Trials, subsequent reviews with various stakeholder consensus groups confirmed that these challenges apply to all DTHs, including those sold as clinical decision support systems. This further analysis identified the seventh challenge:

B7) **Lack of mature business models**: it is still unclear what is the best way to commercialise DTH technologies; in addition to how to sell to end-users, there is an untapped market for Original Equipment Manufacturers (OEM), companies that sell data and software that other companies can use to develop and sell DTH solutions.

To the authors' knowledge, such a systematic analysis is not available for data-driven models, but many of these barriers are common. The lack of advanced models in data-driven models emerges from the lack of high-quality annotated data required to build such models. Besides the lack of validation collections for data-driven models, where building and validating the model requires a large volume of data, data access and data management infrastructures are major problems closely related to model building.

This position paper introduces the Virtual Human Twin (VHT) concept. While there is an ongoing consensus process to reach a robust definition for the VHT, led by the EDITH support action, its general features are defined. Whereas a Digital Twin in Healthcare is a vertical technology aimed to solve a particular clinical problem, the Virtual Human Twin will be a distributed, collaborative infrastructure, a collection of technologies and resources (data, models) that enables it, and a collection of Standard Operating Procedures (SOP) that regulate its use. The Virtual Human Twin infrastructure aims to help address all seven challenges listed above and facilitate researchers and developers in academia, public organisations, and industry to produce new and interoperable DTH solutions. Two essential features must be stressed from the outset. The first is that the Virtual Human Twin is **not** a model able to predict every aspect of human pathophysiology; it is an infrastructure which will allow the accumulation and interconnection of all the quantitative data and knowledge on human pathophysiology. The second is that the VHT will preferentially host **quantitative** (data that can be counted or measured in numerical values), **individual** data, and from **humans**. Preferential means that qualitative data, cohort averages, and animal data will be accepted only as surrogates of quantitative, individual human data when these are impossible to obtain.

If the data the VHT hosts are quantitative, the models in the VHT should also express an expected causal relation between quantities. This excludes, for example, all data-driven models that make predictions from qualitative data, such as clinical reports.

## IV. From the Challenges to the Specifications

Analysing the challenges and leveraging the budding ecosystem allows for identifying several essential (technical) developments for the realisation of a Virtual Human Twin, including but not limited to interoperability, computability, and health data integration. Any VHT blueprint should incorporate regulatory, ethical, legal and social dimensions and specific actions addressing stakeholder needs, implementation, and (clinical/industrial) uptake challenges.

### A. Define the use case: DTH vs VHT

Each Digital Twin in Healthcare is developed with a specific clinical use in mind. But the Virtual Human Twin is not a gigantic digital twin: the VHT is an infrastructure that makes it easier to develop and validate digital twins. This also includes reusing existing models to develop new, more complex models (because of a need to account for multiscale or multisystem processes).

### B. Development: multiscale & multisystem models (B1)

Some biological systems have spatiotemporal characteristics that make scale separation difficult [13], [14]. Also, biological systems are **entangled** in that each sub-system is affected by the state of many other sub-systems. Not surprisingly, the low-hanging fruits for DTH are all problems that can be represented with acceptable accuracy with single-scale, single-subsystem models. Development can be based on curated datasets defined a priori in such cases. But even in those cases, sometimes the need for multiscale and multisystem modelling emerges from the necessity to expose as an input variable a quantity defined at a space-time scale and/or within a physiological sub-system different from that in which the model is defined. However, developing and validating multiscale/multisystem models is challenging, especially regarding the orchestration of multiple single-scale, single-system models. Thus, the Virtual Human Twin should facilitate the development and sharing of such orchestrations and the collection and reuse of the data required to build and validate the models across all different space-time scales they cover and capture the inter-subject variability.

### C. Validation: the data credibility dimension (B2)

Whether the VHT relies on mechanistic models, is driven by AI technology, or integrates both approaches, data will be key. The validation of models with an expected high level of predictive accuracy is challenging. Classic validation approaches assume that the measurement errors affecting the observational data used to validate the model are at least one order of magnitude smaller than the model's prediction errors. But this requires that both the data used as input for the model and those used to validate the model have a very high level of credibility, which means a detailed determination of all measurement errors affecting them. Thus, the Virtual Human Twin should facilitate annotating all data with information on their credibility.

### D. Regulatory: public validation data collection (B3)

Whether they are sold in a country as clinical decision support systems or as medical product (drugs or medical devices) development tools, DTHs need to be reviewed by the regulatory authority competent for that country. For clinical use, the marketing authorisation must follow a regulatory pathway known as *software as a medical device*. It is convenient to pursue a qualification opinion to be used in developing new medical products, which confirms that a methodology is suitable to produce specific evidence in the regulatory process. In both cases, the most challenging part for *in silico* methodologies is the **credibility assessment**, which aims to quantify the accuracy and reliability of the prediction provided by the DTH. Current standards (e.g., ASME VV-40:2018) suggest a one-off credibility assessment is possible for knowledge-driven models. For data-driven models, the topic is still being debated. Still, the predominant orientation is to frame the credibility assessment of data-driven models in the so-called total product lifecycle (TPLC) regulatory approach, where manufacturers commit to transparency and real-world performance monitoring principles. This approach would accommodate the need to retrain machine learning models when new data becomes available.

Whether knowledge-driven or data-driven, evaluating the predictive accuracy of a DTH requires high-quality experimental data and assessing the predictive reliability requires such data to be collected in the most disparate conditions. Unfortunately, producing such extensive, high-quality multidimensional data collections is challenging. Predictive models are used to estimate quantities that are difficult to measure experimentally, so by definition, the generation of these validation data collections is a complex and expensive task.

It is common to produce such data collections to validate a single DTH, typically by the same research group that developed the DTH itself. However, it would dramatically accelerate the development of DTHs if such collections were made publicly available. This would allow groups that do not have the necessary experimental capabilities to validate their DTHs, providing an independent benchmark against which the predictive accuracy of DTHs with the same CoU can be quantified and compared.

### E. Information: a unique information space for all stakeholders (B4)

While validation and regulatory assessment impose a narrowly defined CoU for each DTH, there are various possible uses in research, clinical practice, and product development for the experimental data used to inform and validate a model and its predictions. But the current reuse of such data is minimal, simply because accessing these data is, in most cases, nearly impossible, if not through a collaboration with the team that produced the data. A unique information space would enable all stakeholders (academic researchers, industries, clinical research organisations, consultants, regulators, policymakers, etc.) to access large volumes of data, enabling the most disparate reuses.

Stakeholders are essential to any ecosystem developing new technologies and creating sustainable added value. Healthcare professionals, patients, citizens, regulators, policymakers, payers, and buyers must interact closely with academia and industry to understand needs and build mutual trust. Additionally, in the case of the Virtual Human Twin, many of these stakeholders will be potential active users of the developed technologies. Hence the VHT blueprint should provide clear answers to elements such as user needs and expectations, accessibility, and user interfacing, in addition to addressing the reuse of data and models.

### F. Scalability: machine learning to speed up mechanistic models (B5)

In general data-driven models require significant computational power during their development (training), but not during their execution. On the contrary, knowledge-driven models present the same computational cost in development and use [15].

Combining the two approaches to modelling sometimes improves computational efficiency. Data-driven methods, particularly artificial intelligence and machine learning (AI/ML) techniques that require extensive data collection to train the predictors can be used in developing DTHs in various ways. For example, they can provide the main predictor when no reliable mechanistic knowledge is available for the phenomenon being modelled. AI can use mechanistic variables that are not measurable as input data. Still, they can also be used as a surrogate model of a knowledge-driven model, which provides the same prediction with comparable accuracy but at a fraction of the computational costs. Last, new approaches like physics-informed machine learning, useful in cases where a portion of the phenomenon being modelled has a reliable quantitative mechanistic explanation in terms of laws of physics, can significantly reduce the computational cost of training a new AI/ML model.

Whether AI/ML methods are used to build DTHs or to speed them up, their development requires extensive data collection; in the first case, the data must be experimental, but when the AI/ML is used to produce a surrogate model of the knowledge-driven predictor, the training set is an extensive collection of inputs-outputs pairs generated by running the knowledge-driven model over a large number of different input sets. Many knowledge-driven models are computationally expensive (which motivates the need for a surrogate model). If this is the case, generating such collections may require huge computational resources, which underlines the importance of finding ways to reuse such collections.

### G. Training: the ultimate teaching tool (B6)

There still needs to be more training and re-training on the development and the appropriate use of DTHs. A few elements of *in silico* medicine are included in the curricula of biomedical engineers, but rarely with the specialist focus required to become developers of DTH technologies. More needs to be done to alphabetise (bio)medical students on the potential and limitations of *in silico* methodologies. As the community works to develop the necessary educational content, the possibility of searching an extensive repository of well-curated data and models could become the ultimate teaching tool in several educational scenarios. Imagine how simpler it would be to explain the intricacies of blood sugar homeostasis if each student could run a virtual human on his laptop, fiddle with the level of the various substances involved, and observe the effect.

### H. Business models: data and models sharing (B7)

Such a complex resources sharing scenario must support non-profit and for-profit business models. The infrastructure should inherently support all fundamental Open Access tenants, including the FAIR principles. But it should also enable and support reward mechanisms. Besides the possibility of selling data and model access to commercial users, the infrastructure should make it simple for those who share precious data to be rewarded with access to special data.

With many DTHs and resources still at relatively low Technology Readiness Levels, substantial R&D investments will remain necessary. However, this by itself will not be sufficient to establish an economic sector without the development of an economic value chain and ecosystem around VHT, with goods and service producers and consumers, prescribers, payers, and supporting players such as certifiers, data or computational resource providers, and consolidated distribution channels. Whilst other innovation-intensive health tech sectors can provide some relevant information regarding viable business models for VHT, the strong dependence on data comes with unique risks and uncertainties. As much still needs to be understood, a thorough study of the market potential and an analysis of the few cases of successful translation of DTH technologies will be vital.

## V. Blueprint for the Virtual Human Twin

The specifications listed above provide the general outline for the blueprint of the Virtual Human Twin. This section provides a high-level description of the essential characteristics of the VHT.

### A. Definition

The Virtual Human Twin can be imagined as an n-dimensional data space, which DTH models constantly crawl[3].

The associated knowledge space could be represented by knowledge graphs linking data and models.

This data space represents the totality of our current collective, quantitative knowledge of the physiology and pathology of *homo sapiens sapiens* (the *homo sapiens* species had various sub-species during its evolution; the term *homo sapiens sapiens* is used to indicate the sub-species currently inhabiting the earth, modern humans). Its primary goal is to simplify DTH models' development, validation, integration, and adoption.

Thus, we propose the following definition: “The Virtual Human Twin is an infrastructure, a collaborative distributed knowledge repository and simulation platform of quantitative human (patho)physiology, designed specifically to accelerate the development, the integration, and adoption of patient-specific predictive computer models as clinical decision support systems or as methodologies for the development and derisking of new medical products”.

### B. The data objects

Within the VHT, the atomic entities are data objects and model objects. Predicted data in the context of this paper is defined as data that is obtained as the result of running an *in silico* model (be it a data-driven one, a knowledge-driven one or a combination of both). Each VHT data object is a digital dataset, stored and annotated according to some basic rules. The dataset must contain quantitative information on human pathophysiology, whether measured or predicted. It must be stored and curated according to the FAIR principles, to be findable, accessible (possibly through authentication and authorisation), interoperable and reusable.

The dataset must be annotated with a minimum set of metadata, including information on the data object type and its position in the data space. The Data Object Type (DOT) is a unique identifier associated with enough information to decide if and to what extent that data object is suitable input for a DTH model. This includes information on the dataset regarding its semantics (what the data mean), its syntax (in which standardised, interoperable formats the dataset is accessible), and its accessibility (how the dataset can be accessed). Eventually, DOTs will be selected from a list of standardised types, possibly organised in a well-structured taxonomy or ontology. But for some time, the list of supported DOTs might be a *folksonomy*, a user-generated way of organising content, which is periodically scrutinised and consolidated into proper ontologies. When a new model is added to the VHT, first one needs to check if, for each input and output of the model, a valid DOT has already been defined; if not, before the model can be published on the VHT, the new DOTs need to be added.

In computer vision and robotics, the pose of an object is the combination of the object's position and orientation. Pose estimation determines a detected object's pose relative to some coordinate system. This information can then be used, for example, to allow a robot to manipulate an object or to avoid moving into the object. The Data Object Pose (DOP) includes all information to define the position of the data object in the VHT six-dimensional reference system and the scale information, such as the grain and range of the dataset [13]. Grain is defined as the larger of (i) the minimum distance (or time span) that can be distinguished by the instrumentation or (ii) as the characteristic distance (or time span) of variation of the smallest (or fastest) feature of interest measured using this instrumentation. Extent is defined as the smaller of (i) the maximum distance (or time span) that the same instrumentation can measure or (ii) as the characteristic distance (or time span) of variation of the largest (or slowest) feature of interest measured using this instrumentation.

The six dimensions of the data space are represented in Figure 1 and defined in the following sections. The concept of grain and range as scale representation applies well to datasets that define the variation of a quantity in space and time. But since we assume by convention, as described above, that also scalar values are associated with a point in the 6D reference system of the VHT, in that case, the grain represents the least significant digit of the measurement/prediction (reproducibility of the measurement, uncertainty of the prediction). In contrast, the range could be used to represent the uncertainty of positioning in space and time for that scalar quantity.

[3] “crawl” is not a technical term, and some may find it confusing. The idea is that model objects are like little insects that crawl the honeycomb of data objects, “eat” some data objects from certain honeycomb cells and “lay” some new data objects in other cells.

Whenever a new DOT is added, it should also be provided with the *transformation functions* required to calculate the DOP for each data object with that DOT. See Figure 1 for a schematic overview of the DOT/DOP and the text below the figure for a more detailed description of the data space and its six dimensions.

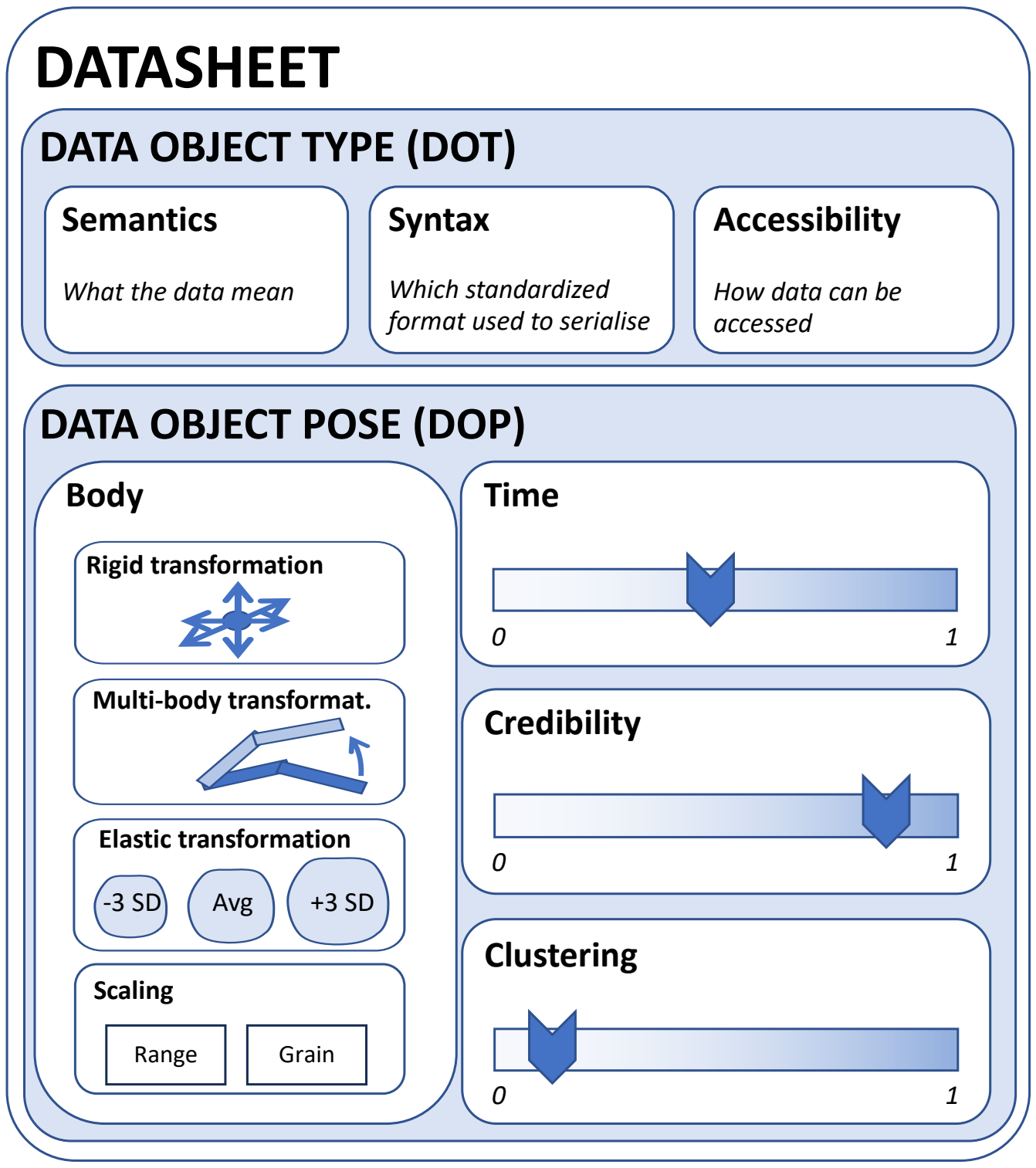

Figure 1. A graphic representation of the data sheet describing the relevant information of a data object.

## C. The model objects

In VHT, model objects are defined as **data space crawlers**. A VHT model requires a finite number of inputs, described in terms of DOTs and DOPs, and produces, upon successful execution, a certain number of outputs, also described in terms of DOTs and DOPs. When a model is active, every time a new data object with the necessary DOT is added to the data space, the VHT model is automatically executed. Its outputs are also added to the data space in the appropriate DOP. Thus, every time we add to the VHT a group of data objects that constitute a valid input for a model object, the dataspace will automatically enrich with new predicted data. This implies that VHT models must execute in batch mode; however, human interaction is still possible using a "person-in-the-middle" paradigm [16]. It should be noted that this "greedy" execution model is only a conceptual representation; in practice, when a new valid input is found in the data space, the model goes into an execution queue, where it will stay until someone provides the necessary computational resources.

Two important technical aspects need to be addressed: remote execution and orchestration. The VHT will run on a single computer cluster with some storage in the simplest scenario. All data objects are stored in this storage, and all model objects execute on the computer cluster. But as soon as we imagine more complex architectures inherent to federated approaches, we might have a situation where the storage that contains the data objects, and the computer that executes the model objects, are not co-located. To ensure maximum flexibility, we can imagine a scenario where both data objects and model objects are portable, the first using data replication services and the second using container architectures. This would allow the creation of a rule-based system that decides case by case if it is better to move the data or the models.

The second issue is model orchestration. As we mentioned above, many problems require the orchestration of multiple models, where the outputs of one model are the inputs of another. In many cases, model orchestration can be formulated exclusively in terms of data flow. Model A reads its inputs and calculates its outputs. Model B reads these outputs as inputs and calculates its outputs. And so on. This type of orchestration can be provided as a by-product of the proposed data space crawler architecture. Essentially model B executes immediately upon the appearance in the data space of the outputs of model A, which model B considers as valid inputs. The complexity of the orchestration topology is not critical since the data flow can handle virtually any topology. The only exceptions are the so-called "strongly coupled" models. These are models where executing the next step in the calculation of model B requires the results of the current step of the calculation of model A. Strongly coupled models run simultaneously, exchanging data as they go through a shared file system or through the computer memory. These models can be orchestrated using specialised libraries [17]. In this case, for the VHT, the whole orchestration would be seen as a single model object.

## D. The six dimensions of the data space

Accessing such complex data space requires some sort of organisational paradigm. So far, we have identified six dimensions along which the data space can be organised. These are Space (Body Height, Body Width, Body Depth), Time, Credibility and Clustering. All this information combined for a given data set amounts to the DOP.

### 1) Body

We assume there is a 3D model of the average human body in the conventional standing pose (VHT Anatomical Template) that is considered the anatomical template on which we spatially map each data object.

Data objects can be defined over 0, 1, 2, or 3 spatial dimensions. For example, the systolic blood pressure of a subject is a 0D data object; how the blood flow velocity varies along the length of an artery is a 1D data object; the distribution of temperature over a region of the skin is a 2D data object; the distribution of bone mineral density in a bone is a 3D data object.

Each data object (except 0D objects) represents the spatial variation of its values using an implicit reference system; so, in a 3D data object, the value corresponding to the coordinates (0, 0, 0) places such value at the origin of this implicit reference system. In addition, each data object is referred to a specific individual and their anthropometry. But to simplify the automatic annotation, the clustering, and other similar operations, it is convenient that each data object is mapped to a conventional anatomical space by posing it with respect to the VHT anatomical template (which provides the DOP).

3D objects can be easily posed in the anatomical space; the bone mineral density distribution of a patient's femur can be posed in the VHT Space region corresponding to the femur of the anatomical template. With some caution, 2D and 1D objects can also be posed with respect to the anatomical template. However, 0D objects do not have an anatomical location. But because all data objects in the VHT must have one, all 0D data objects are, by convention, mapped on a 3D point located in a conventional point in the anatomical space. So, for example, the systolic blood pressure value could be posed at the centre of the heart region in the anatomical template or in the arm region where the sphygmomanometer was applied.

Two spatial transformation functions should be available for a given DOT: rigid roto-translation and elastic registration. The rigid roto-translation can consider the data object as a single rigid body or a kinematic chain depending on the data type. This function calculates the spatial transformation required to align the data object to the VHT anatomical template and is stored in the DOP of the data object. For example, the elastic registration transformation function is used when we compute the average geometry for a sub-population (which corresponds to a coordinate on the clustering axis).

Among the essential metadata for each DOT, one must include the spatial range and grain of the data object [13], which facilitates the definition of the spatial scale in which the data object is defined.

It should be noted that this grounding of the data to the anatomy poses some challenges. A major one is how to handle datasets that refer to multiple anatomical locations, for example, the recordings of a multi-lead electrocardiogram. In such cases, one could position the dataset in correspondence with the heart centre or at the centre of the chest region. Or, if the anatomical location of each lead is available, one could decompose the dataset into multiple data objects, one for each channel, and place them at the anatomical location of their lead. In this case, metadata should retain the information that the same experiment produced all those measurements.

*2) Time*

The data object is positioned in the Time axis according to the data collection date. Depending on the needs, this information can be normalised onto two additional representations. If the subject's birthdate is known, the data object can be mapped onto the Age representation of the time axis. If the subject's death date is known, the lifespan representation can also be used, spanning from zero (birth) to one (death). A scaling factor vector is calculated for each time coordinate for time-varying data objects. Since the time of birth is not generally available information, we will assume that all subjects born on a given day were born at 12:00 (noon) because sixty per cent of babies are born during the day, between 6 A.M. and 6 P.M.

Among the essential metadata for each DOT, one must include the temporal range and the grain of the data object, which makes it possible to define the time scale in which the data object is defined.

*3) Credibility*

When a new data object is added, it is placed at the lowest level of credibility (non-qualified data). The data owner can submit a data object to the credibility transformation function. The higher the credibility of a data object, the higher its value.

Depending on the level of credibility that the owner is requesting, the application must be informed by a smaller or greater amount of information that captures the provenance, the quality, the metrological properties (or computational credibility properties if the data are computed), and the certifications of the instrumentation/software used. For high levels of credibility, the request might be evaluated by a panel of experts, possibly in coordination with regulatory agencies. An important aspect to consider is the propagation of uncertainty. Uncertainty quantification can be used to evaluate how the uncertainty of input data affects a model's predictions. Still, it is necessary to define also how to express the credibility of models built as orchestration of other models.

*4) Clustering*

Each DOT must include among its transformation functions an averaging function that enables clustering. For data objects defined in space, this is typically an elastic registration function; for time-varying objects, it might involve a synchronisation function; for data objects not defined in space-time, these are more properly averaging functions in the statistical sense.

When added to the VHT, each data object is placed at clustering k = 0 (no clustering). To ensure irreversible anonymisation, the metadata includes a unique data object ID and a unique PatientID, not associated with the individual identity. Where necessary, a LocalPatientID can be used to support pseudo-anonymisation schemes.

All data objects are automatically added to one default cluster: *homo sapiens* (k = 1). Specific research projects may calculate other sets stored with enough metadata to inform the number of groups and the criteria used for clustering. This means that on the Clustering axis, there might be in the same coordinate multiple data objects for the same DOT type, each obtained with different clustering criteria. So, for example, under k = 0.5, we could have a male-female clustering, a healthy-diseased clustering, or a clustering over/below 40 years of age. More complex approaches to clustering might be explored with knowledge graphs.

*5) An example: generation of the anatomical template*

We use the hypothetical generation of the VHT Anatomical Template to illustrate how these six dimensions are defined. Let us imagine having a large collection of 3D body scans of humans of all ages, genders, etc. In theory, all scans were taken with the subject in the same conventional pose (standing with the feet slightly apart, arms along the sides with the palms forward).

Each dataset is expressed with respect to an implicit reference system specific to the type of scanner used. We position the data object on the time axis in correspondence to the date when the scan was performed. If available, the subject's birthdate lets us calculate the age, and the subject's death date lets us calculate the time normalised over the life span.

Assuming the scans were all performed with fully certified 3D scanners, we place all datasets at 1 on the credibility axis (which ranges from 0 for non-qualified data to 1 for fully certified measured data).

Since each dataset refers to an individual, we place all of them at 0 on the clustering axis (the degree of clustering $k$ is defined as $k = 1/x$, where $x$ is the number of clusters: the *homo sapiens sapiens* cluster has $k = 1$; male-female clustering has $k = 0.5$; and individual datasets have $k = 0$, assuming an infinite number of human beings).

If we now select all datasets for individuals of a certain age, we can perform some spatial normalisations. The first normalisation operation assumes the body is a rigid object. We define an anatomical reference system (e.g., origin in the projection of the centre of mass on the floor, X oriented from posterior to anterior, Y from medial to lateral, and Z from feet to head) and calculate for each dataset the rigid transformation so that they are all aligned to the anatomical reference system. The second normalisation operation assumes the body is a kinematic chain, e.g., a set of rigid bodies articulated through idealised joints. We define in the anatomical reference system an ideal body posture. Then we calculate the multi-body rigid transformation for each dataset that aligns each scan to this ideal body posture. The third and last spatial normalisation assumes the body is an elastic object. We use statistical atlas techniques [18] to calculate for each time point the average body shape and then calculate the transformation of each dataset to this average body shape. The vector of average body shapes at different ages is the VHT anatomical template. Each new VHT data object must be posed to this anatomical template.

## VI. Ecosystem

The Virtual Human Twin ecosystem could be imagined as in figure 2.

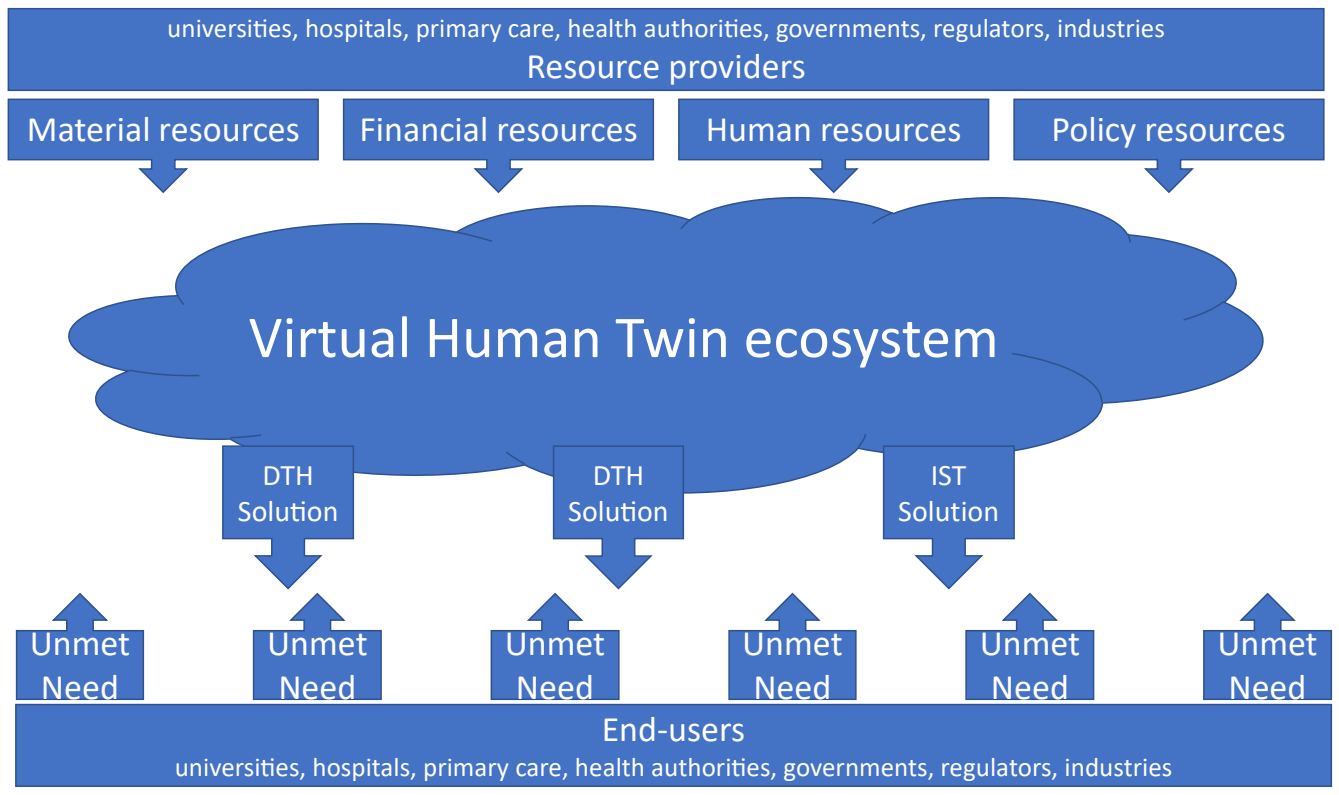

Figure 2. A graphic representation of the ecosystem within which the VHT will develop. DTH: Digital Twin in Healthcare; IST: *in silico* trial.

At the base, all prospective end-users express unmet needs that could be satisfied, at least in principle, with a Digital Twin in Healthcare or an *In Silico* Trial technology. On the top are all those institutions that can contribute material, financial, human and policy resources to developing such technologies. The Virtual Human Twin ecosystem should be a broker between these two groups, leveraging a European infrastructure that facilitates and speeds up such solutions' development and validation. Resources will be published on the VHT by individual researchers and funded consortia; individual researchers and consortia will reuse such resources for their own research purposes.

The distributed nature of the VHT infrastructure will also be essential for the ecosystem's development. Some of the data and models will be bounded to exist only on specific computers/networks because of a variety of reasons, including legal constraints (e.g., privacy laws in clinical data), technical limitations (e.g., particular models can execute only on supercomputers), or commercial conditions (e.g., detailed data or models have regulated access policies). As, at least in principle, all data and models can be seamlessly migrated from one computer to another, the VHT infrastructure should offer sufficient flexibility to support most use policies.

We do not imagine this ecosystem and its infrastructural elements as a fixed, permanent organisation. On the contrary, we imagine it as a dynamic, constantly changing environment. It could start as a strongly pre-competitive ecosystem, strongly subsidised through public research and innovation funding; early participants would trade value mainly through barter mechanisms. As it matures, the value could be exchanged using some form of internal transferable credits, which could also be acquired with cash, which would be used to make the ecosystem at least partially self-sustainable; we would expect public and private investors such as the European Investment Bank to play a significant role in this phase. Another asset that can contribute to the self-sustainability of the ecosystem is the exploitation of certified benchmark datasets and models that would allow a fair, independent, and quantitative comparison of DTH and IST solutions. Eventually, most pre-competitive actors would be replaced by commercial operators. The ecosystem could become entirely immaterial (collection of commercial services glued together by standards and policies), or it could preserve the need for a central entity that could take the form of a public-private partnership.

## VII. From the Blueprint Back to the Challenges

How can this collaborative and distributed knowledge repository and simulation platform, for which we provided a preliminary blueprint, help address the challenges to adopting the DTH technologies?

While a few multiscale problems impose strongly coupled approaches [17], most multiscale/multisystem modelling problems can be formulated primarily in terms of data flow, meaning that these multiscale models are simply orchestrations of the data that flow from one single-scale model to another. The VHT would facilitate the development of these single-scale models if the data representing the phenomenon at a lower or higher scale than the one of interest were already available within the VHT. One researcher could develop a model that takes as input some molecular biomarkers and predicts the disease progression at the cellular scale. Another could develop a model that predicts the clinical symptoms at the organ scale if it is known how the phenomenon manifests at the cellular scale. A third researcher could compose these two models into a multiscale orchestration that can predict the organ-scale symptoms from some molecular biomarkers.

The critical element of such a multiscale model would be the relational model [19], [20], which transforms the data from scale A to scale B and back (typically homogenisation /particularisation algorithms). Each research group could focus on developing and validating models at a single scale or composing data at multiple scales into a multiscale simulation. The same argument applies to multi-system models.

Historically, research groups tend to specialise in experimental or computational methods, making it quite rare to find a group with both skill sets required to address the problem of model validation. Publications are rarely an effective conduit, as much information needed to use experimental data for model validation is omitted from experimental publications. The VHT ensures that the experimental data are carefully annotated with all information necessary to use an experimental dataset to validate a model. Furthermore, it offers a digital agora where groups can directly or indirectly collaborate, e.g., when an experimentalist sees some model's prediction waiting for validation, they can develop an experiment designed to validate such predictions and vice versa. In general, the Credibility axis will enforce a quality assurance culture, where every data object, measured or predicted, is qualified in terms of credibility at the highest possible level.

The availability of validation collections not necessarily produced by the same researchers who developed the model will ensure a source of independent and unbiased comparison between competing models. Once these validation collections are appropriately certified, they could provide a pivotal element to simplify the regulatory process for DTH and *In silico* Trials. One exciting approach to the production of such collections is that of the modelling challenges: there are examples in the fields of biomechanics [21], radiomics [22], [23] and in epidemiology [24], [25]. Another is to orient research funding to the production of such collections.

Methodologically complex approaches such as DTH tend to favour the circulation of information within communities composed around the methodology rather than the research question. While it might be helpful in some cases for modellers to talk to each other, multidisciplinary research domains such as this can flourish only when all types of expertise (clinical, pre-clinical, technological, etc.) are part of the debate. The VHT will favour a re-composition of research communities around research questions and unmet medical needs rather than around methodologies. A cardiologist will access the VHT to find the distribution of cardiac ejection fraction values over a specific population; a pre-clinical researcher to compare the contractility of mouse cardiomyocytes with human ones; a modeller to have their model validated against experimental data. The same argument applies along the entire innovation pipeline, linking academics, industrialists, regulators, payers, healthcare providers, etc. A robust communication space attached to the VHT will ensure that all stakeholders are adequately informed and networked and that their needs and requirements have been considered.

The scalability of DTH will improve dramatically as pre-computed data are made available, and complex multiscale/multisystem simulations can be reduced to orchestration and some scale transformations. Also, every new model added to the VHT will be run on all data available that form a suitable input for such a model according to the DOP. These collections of model predictions are perfect for generating surrogate models that provide the same predictions at a fraction of the computational cost. And the pooling of data collected by various groups might be the only solution for training some AI predictors targeting challenging problems where mechanistic knowledge is still scarce.

The vast collection of data and models is the perfect educational tool to support any curriculum related to *In silico* Medicine.

As a general design principle for the VHT, while most metadata will be publicly available, the data objects might require authentication and/or authorisation to be accessed, a possibility also recognised in the definition of the FAIR principles. This will make the VHT a great socioeconomic experiment where economists and entrepreneurs could explore new business models and strategies to reward the providers of data and models and self-sustain the infrastructure after a development phase where public funding will be necessary.

Last, this paper did not address the need for all data and models exposed through the VHT to comply with the national and supranational ethical and legal constraints. We did not discuss this not because it is unimportant but because how exactly the VHT should address this aspect still needs to be discussed.

This position paper does not address several critical aspects, such as privacy issues, business models, sustainability, etc. Each of these topics is being addressed as part of the consensus process run by the EDITH coordination and support action and will be addressed in the research roadmap this initiative plans to publish in 2024 (a preliminary draft is already available[4]).

In conclusion, we believe the creation of a collaborative distributed knowledge repository and simulation platform of quantitative human pathophysiology described in this position paper could dramatically accelerate the development and adoption of patient-specific predictive computer models to be used as clinical decision support systems or as a tool to refine, reduce and replace *in vitro*, animal, and human experimentation for the assessment of new medical products.

## Acknowledgement

The European Commission partially supported this study through the H2020 project "*In silico* World: Lowering barriers to ubiquitous adoption of *In silico* Trials" (topic SC1-DTH-06-2020, grant ID 101016503) and the Horizon Europe

[4] https://zenodo.org/record/8070381

## REFERENCES

[1] «Unmet health care needs statistics». https://ec.europa.eu/eurostat/statistics-explained/index.php?title=Unmet_health_care_needs_statistics (consulted on Dec 24th, 2022).

[2] P. Hunter, P. Robbins, e D. Noble, «The IUPS human Physiome Project», Pflugers Arch, vol. 445, fasc. 1, pp. 1–9, ott. 2002, doi: 10.1007/s00424-002-0890-1.

[3] «The Human Genome Project», Genome.gov. https://www.genome.gov/human-genome-project (consulted on Dec 24th, 2022).

[4] J. W. Fenner et al., «The EuroPhysiome, STEP and a roadmap for the virtual physiological human», Philos Trans A Math Phys Eng Sci, vol. 366, fasc. 1878, pp. 2979–2999, set. 2008, doi: 10.1098/rsta.2008.0089.

[5] Discipulus Consortium, «Digital Patient Roadmap», ott. 2013. [Online]. Available at: http://www.vph-institute.org/upload/discipulus-digital-patient-research-roadmap_5270f44c03856.pdf

[6] C. A. Taylor, T. A. Fonte, e J. K. Min, «Computational fluid dynamics applied to cardiac computed tomography for noninvasive quantification of fractional flow reserve: scientific basis», J Am Coll Cardiol, vol. 61, fasc. 22, pp. 2233–2241, giu. 2013, doi: 10.1016/j.jacc.2012.11.083.

[7] STEP Consortium, «Seeding the EuroPhysiome: a roadmap to the virtual physiological human», VPH Institute, Brussels, Belgium, 2007. [Online]. Available at http://www.vph-institute.org/upload/step-vph-roadmap-printed-3_5192459539f3c.pdf

[8] P. Hunter et al., «A vision and strategy for the virtual physiological human in 2010 and beyond», Philos Trans A Math Phys Eng Sci, vol. 368, fasc. 1920, pp. 2595–2614, giu. 2010, doi: 10.1098/rsta.2010.0048.

[9] M. Viceconti e P. Hunter, «The Virtual Physiological Human: Ten Years After», Annu Rev Biomed Eng, vol. 18, pp. 103–23, lug. 2016, doi: 10.1146/annurev-bioeng-110915-114742.

[10] M. Viceconti, A. Henney, e E. Morley-Fletcher, «*In silico* clinical trials: how computer simulation will transform the biomedical industry», International Journal of Clinical Trials, vol. 3, p. 37, mag. 2016, doi: 10.18203/2349-3259.ijct20161408.

[11] M. W. Grieves, «Virtually Intelligent Product Systems: Digital and Physical Twins», in Complex Systems Engineering: Theory and Practice, vol. Volume 256, American Institute of Aeronautics and Astronautics, Inc., 2019, pp. 175–200. doi: 10.2514/5.9781624105654.0175.0200.

[12] T. G. Buchman, «The digital patient: predicting physiologic dynamics with mathematical models», Crit Care Med, vol. 37, fasc. 3, pp. 1167–1168, mar. 2009, doi: 10.1097/CCM.0b013e3181987bbc.

[13] P. Bhattacharya, Q. Li, D. Lacroix, V. Kadirkamanathan, e M. Viceconti, «A systematic approach to the scale separation problem in the development of multiscale models», PLoS One, vol. 16, fasc. 5, p. e0251297, 2021, doi: 10.1371/journal.pone.0251297.

[14] B. de Melo Quintela, S. Hervas-Raluy, J. M. Garcia-Aznar, D. Walker, K. Y. Wertheim, e M. Viceconti, «A theoretical analysis of the scale separation in a model to predict solid tumour growth», J Theor Biol, vol. 547, p. 111173, ago. 2022, doi: 10.1016/j.jtbi.2022.111173.

[15] Geris Liesbet, Rousseau Cécile F., Noailly Jerome, Afshari Payman, Auffret Michaël, Chu Wen-Yang, De Cunha-Burgman Martha, Lafon Yoann, Marchal Thierry, Palmer Mark, Voisin Emmanuelle M., Viceconti Marco, Hoekstra Alfons G., Johnston Gordon, Alessandrello Rossana, Bril Antoine, Dall'Ara Enrico, Kulesza Alexander, Lesage Raphaëlle, … Waterkeyn Alicia. (2022). WHITE PAPER: the role of Artificial Intelligence within in silico medicine. Zenodo. https://doi.org/10.5281/zenodo.8064147

[16] M. Yetisgen-Yildiz, I. Solti, e F. Xia, «Using Amazon's Mechanical Turk for Annotating Medical Named Entities», AMIA Annu Symp Proc, vol. 2010, p. 1316, 2010.

[17] J. Borgdorff et al., «Distributed multiscale computing with MUSCLE 2, the Multiscale Coupling Library and Environment», Journal of Computational Science, vol. 5, fasc. 5, pp. 719–731, set. 2014, doi: 10.1016/j.jocs.2014.04.004.

[18] A. Sotiras, C. Davatzikos, e N. Paragios, «Deformable medical image registration: a survey», IEEE Trans Med Imaging, vol. 32, fasc. 7, pp. 1153–1190, lug. 2013, doi: 10.1109/TMI.2013.2265603.

[19] G. Stamatakos et al., «Computational Horizons In Cancer (CHIC): Developing Meta- and Hyper-Multiscale Models and Repositories for *In silico* Oncology - a Brief Technical Outline of the Project», Proc 2014 6th Int Adv Res Workshop *In silico* Oncol Cancer Investig (2014), vol. 2014, nov. 2014, doi: 10.1109/iarwisoci.2014.7034630.

[20] D. Tartarini, K. Duan, N. Gruel, D. Testi, D. Walker, e M. Viceconti, «The VPH Hypermodelling framework for cancer multiscale models in the clinical practice», in Proceedings of the 2014 6th International Advanced Research Workshop on *In silico* Oncology and Cancer Investigation - The CHIC Project Workshop (IARWISOCI), Athens, Greece, nov. 2014, pp. 1–4. doi: 10.1109/IARWISOCI.2014.7034642.

[21] B. J. Fregly et al., «Grand challenge competition to predict in vivo knee loads», J Orthop Res, vol. 30, fasc. 4, pp. 503–513, apr. 2012, doi: 10.1002/jor.22023.

[22] Y. Sun et al., «Multi-Site Infant Brain Segmentation Algorithms: The iSeg-2019 Challenge», IEEE Trans Med Imaging, vol. 40, fasc. 5, pp. 1363–1376, mag. 2021, doi: 10.1109/TMI.2021.3055428.

[23] J. Hirvasniemi et al., «The KNee OsteoArthritis Prediction (KNOAP2020) Challenge: An image analysis challenge to predict incident symptomatic radiographic knee osteoarthritis from MRI and X-ray images», Osteoarthritis Cartilage, pp. S1063-4584(22)00864–0, ott. 2022, doi: 10.1016/j.joca.2022.10.001.

[24] M. Ajelli et al., «The RAPIDD Ebola forecasting challenge: Model description and synthetic data generation», Epidemics, vol. 22, pp. 3–12, mar. 2018, doi: 10.1016/j.epidem.2017.09.001.

[25] P. Ezanno et al., «The African swine fever modelling challenge: Model comparison and lessons learnt», Epidemics, vol. 40, p. 100615, set. 2022, doi: 10.1016/j.epidem.2022.100615.